# Optical Tomographic Imaging for Breast Cancer Detection

Wenxiang Cong, Xavier Intes, Ge Wang

Biomedical Imaging Center, Rensselaer Polytechnic Institute, Troy, New York 12180

*Abstract*— **Diffuse optical breast imaging utilizes near-infrared (NIR) light propagation through tissues to assess the optical properties of tissue for the identification of abnormal tissue. This optical imaging approach is sensitive, cost-effective, and does not involve any ionizing radiation. However, the image reconstruction of diffuse optical tomography (DOT) is a nonlinear inverse problem and suffers from severe ill-posedness, especially in the cases of strong noise and incomplete data. In this paper, a novel image reconstruction method is proposed for the detection of breast cancer. This method split the image reconstruction problem into the localization of abnormal tissues and quantification of absorption variations. The localization of abnormal tissues is performed based on a new well-posed optimization model, which can be solved via differential evolution optimization method to achieve a stable image reconstruction. The quantification of abnormal absorption variations is then determined in localized regions of relatively small extents, which are potentially tumors. Consequently, the number of unknown absorption variables can be greatly reduced to overcome the underdetermined nature of diffuse optical tomography (DOT), allowing for accurate and stable reconstruction of the abnormal absorption variations in the breast. Numerical simulation experiments show that the image reconstruction method is stable and accurate for the identification of abnormal tissues, and robust against measurement noise of data.**

*Index Terms*—**Diffuse optical tomography (DOT), breast imaging, image reconstruction, differential evolution (DE).**

## I. INTRODUCTION

B reast cancer is a major health problem. If this disease is treated at an early-stage using current therapies, patient prognosis will be improved significantly. Currently, multiple methods are available for breast cancer screening and diagnosis. The x-ray mammography is an effective screening technique. However, the breast consists of tissues with similar densities and has similar attenuation coefficients. The appearance of cancer on mammograms has a substantial similarity to that of normal tissues. The x-ray mammography is less sensitive in dense breast [1]. X-rays are also a type of ionizing radiation, which bears a risk of cancer induction. Magnetic Resonance Imaging (MRI) is another effective and sensitive technique to detect cancer in dense tissues. It has a very high negative predictive value, which helps distinguish benign tumors and malignant tumors, and decrease the possibility of false negative diagnosis. A major disadvantage of MRI is much more expensive and time-consuming than other breast-diagnostic methods. Ultrasonography (US) is delineate cysts, benign and malignant masses. But it is limited by poor soft tissue contrast, inherent speckle noise, and strong operator dependence [2]. The optical molecular imaging modality uses exogenous fluorescent probes as additional contrast agents that target molecules relevant to breast cancer [3]. The use of fluorescent probes has a potential in early breast cancer detection but the effectiveness of fluorescence imaging relies on the functions of the probes [4]. Molecular contrast probes on specific tumor receptors or tumor associated enzymes are also under development.

Diffuse optical tomography (DOT) was introduced the early 1990s. DOT uses near-infrared light transmission and intrinsic breast tissue contrast for the detection and characterization of abnormal tissue [5]. It is sensitive, cost-effective, and does not involve any ionizing radiation, and has a high sensitivity due to the rich optical absorption contrast. Owing to the relatively low absorption of hemoglobin, water and lipid at wavelengths of 650–1000 nm, near-infrared (NIR) light can transmit through several centimeters of biological tissue with an adequate signal-to-noise ratio for breast tomographic imaging [6]. Furthermore, in this spectral range, oxy-hemoglobin and deoxy-hemoglobin and lipids predominantly affect the absorptive properties of the breast tissue. There is a difference in total hemoglobin concentration levels between benign and malignant breast lesions, so it was a useful indicator for distinguishing between benign and malignant breast lesions [7]. By combining images reconstructed by DOT at various wavelengths, concentrations of oxy- and deoxy hemoglobin, and water can be determined to reveal tumors from background tissue [8].

However, NIR light is strong scattering in biological tissues, the measurable quantity can only be collected on a partial external surface of an object, and light fields obtained from different excitation sources are highly correlated. The resultant volumetric reconstruction is a nonlinear inverse problem and suffers from severe ill-posedness exaggerated by data noise and measurement incompleteness. The efficient, stable, and accurate image reconstruction is the most challenging imaging tasks. The Born and Rytov approximation are popular linearization methods [9], which simplify the inverse problem as a linear system to describe the relationship between tissue absorption coefficients and boundary measurement. The regularization or stabilization techniques are applied to obtain physically realistic results. In the case of sparse distribution of optical parameters, the total variation (TV) regularization [10] is more efficient than generic regularization. Sparse image reconstruction with the $l_1$ norm regularization is also helpful to achieve a stable image recovery from incomplete and inaccurate measurements [11, 12]. However, these methods cannot fundamentally solve the ill-posed problem of the diffuse optical tomography, usually resulting in aberrant image reconstruction in the presence of measurement noise.

*This work was supported by the National Institutes of Health Grant NIH/NIBIB R01 EB019443

In this paper, we propose a new diffuse optical tomographic imaging method for the detection of breast cancer. This method divides the image reconstruction problem into the localization of abnormal tissues and the quantification of absorption variations. The localization of abnormal tissues is performed based on a well-posed optimization model, which can be solved using differential evolution global optimization method. The quantification of abnormal absorption variations is then determined in localized regions of relatively small extents, which are potentially tumors. This method is robust against measurement noise of data, and produces accurate and stable image reconstruction of abnormal absorption variations in the breast.

## II. IMAGE RECONSTRUCTION

### A. Imaging system design

The optical breast imaging system consists of an array of photon detectors, a light source plate, and system frame. An array (64×64) of avalanche photodiode (APD C12402-04, Hamamatsu) is assembled into a 2D detector array to measure optical signals [13]. APD has excellent photon-counting capability for detecting extremely weak light at the photon counting level, a fine time resolution, and a wide spectral response range. Sixty-four optical fibers are arranged into 8 rows by 8 columns, and mounted on the transparent plate. The breast can be slightly compressed between the two parallel glass plates. A NIR laser (725nm-875nm) is connected with a fiber channel to excite the breast. The light source and detectors are placed on the opposite sides of the breast, as shown in Figure 1. The source and detectors can be rotated to change the excitation-detection arrangement for sufficient

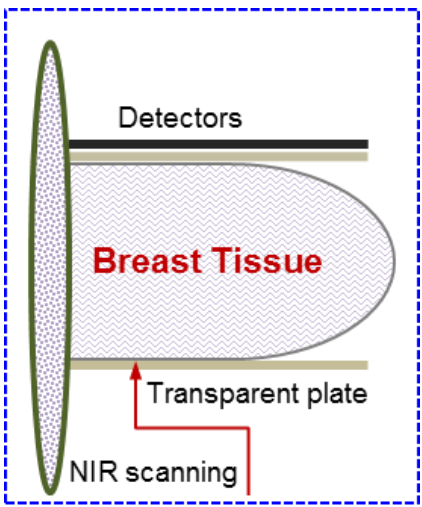

. **Figure 1. Schematic diagram of breast optical imaging system**

signal acquisition. During an optical scanning, the breast is sequentially illuminated by switching the near-infrared laser to different fiber channels. Light signals emanating from the breast is measured by the detectors for each source position. The measured dataset for every source-detector pair is used for image reconstruction of absorption variations. The system is calibrated and optimized for high-performance imaging. The imaging system can also be adapted to time-resolved/frequency domain imaging mode.

### B. Physical model

Optical imaging utilizes light propagating through tissue for assessment of optical properties. The propagation of light in biological tissues is a complex process involving both absorption and scattering of light. The light propagation model describes the interaction of photons with tissues and predicts a light intensity distribution in the breast and on its external surface. By minimizing the discrepancy between the predicted data and measured data, an image reconstruction algorithm can recover the spatial distribution of intrinsic optical tissue properties. The radiative transfer equation (RTE) accurately models the photon propagation in biological tissues [14, 15]. Due to the extensive computational cost, it is difficult to directly apply RTE for biomedical tomographic imaging. The diffusion approximation (DA) model to RTE is widely used to describe the NIR light transport in biological tissues with an adequate accuracy in strongly scattering and weakly absorbing objects, and a high computational efficiency [14]. The steady-state DA model is expressed as follows:

$$-\nabla \cdot \left[ D(\mathbf{r}) \nabla \Phi(\mathbf{r}) \right] + \mu_a(\mathbf{r}) \Phi(\mathbf{r}) = S(\mathbf{r}), \quad \mathbf{r} \in \Omega \quad (1)$$

where $\mathbf{r}$ is a positional vector, $\Phi(\mathbf{r})$ is the NIR photon fluence rate, $S(\mathbf{r})$ is the NIR source, $\mu_a(\mathbf{r})$ is the absorption coefficient, $D(\mathbf{r})$ is the diffusion coefficient defined by $D(\mathbf{r}) = \left[ 3(\mu_a(\mathbf{r}) + \mu_s'(\mathbf{r})) \right]$, $\mu_s'(\mathbf{r})$ is the reduced scattering coefficient, and $\Omega \subset R^3$ a region of interest (ROI) in the object. If no photon travels across the boundary $\partial \Omega$ into the tissue domain $\Omega$, DA is constrained by the Robin boundary condition [16]:

$$\Phi(\mathbf{r}) + 2\alpha D(\mathbf{r})(\upsilon \cdot \nabla \Phi(\mathbf{r})) = 0, \quad \mathbf{r} \in \partial \Omega, \quad (2)$$

where $\upsilon$ is an outward unit normal vector on $\partial \Omega$, and $\alpha$ the boundary mismatch factor between the tissue with a refractive index $n$ and air, which can be approximated by $\alpha = (1 + \gamma)/(1 - \gamma)$ with $\gamma = -1.44 n^{-2} + 0.71 n^{-1} + 0.67 + 0.06n$ [16]. The measurable photon fluence on the surface of the object can be expressed as,

$$m(\mathbf{r}) = -D(\mathbf{r})(\mathbf{v} \cdot \nabla \Phi(\mathbf{r})), \quad \mathbf{r} \in \partial \Omega. \quad (3)$$

Breast tissue consists mainly of light elements with low atomic numbers, and its elemental composition is nearly uniform with little density variation. Hence, the spatial variation of scattering coefficients in breast tissues is slow [11]. Hence, in this context, we only focus on the absorption contrast since angiogenesis in cancer mainly causes a change in absorption properties of hemoglobin. Accordingly, we can decompose the absorption coefficients into $\mu_a(\mathbf{r}) = \mu_a^0(\mathbf{r}) + \delta\mu_a(\mathbf{r})$, where $\mu_a^0(\mathbf{r})$ is the absorption coefficient of normal breast tissue, and $\delta\mu_a(\mathbf{r})$ denotes the abnormal absorption variations in tissues, since malignant tissues show higher levels of absorption compared to healthy tissues. Hence, we can rewrite Eq. (1) as follows:

$$-\nabla \cdot \left[ D(\mathbf{r}) \nabla \Phi(\mathbf{r}) \right] + \mu_a^0(\mathbf{r})\Phi(\mathbf{r}) + \delta\mu_a(\mathbf{r})\Phi(\mathbf{r}) = S(\mathbf{r}), \mathbf{r} \in \Omega \quad (4)$$

Eq. (4) is the governing imaging model for the image reconstruction of abnormal tissues.

### C. Localization of abnormal absorptions

The geometrical shape of the compressive breast can be accurately established via laser scanning (http://www.nextengine.com). From the geometrical shape of the compressive breast, a finite-element mesh is generated using the computer graphic techniques (Amira software). Key reference points are marked on breast. Accordingly, the markers also appear on the measured photon density image. Thus, the one-to-one relationship between all pixels on the measured photon density image and the coordinates of the finite elements model can be established for the image registration. Based on finite element model, differential equations (2) and (4) can be discretized into a matrix equation with respect to the photon fluence rate $\Phi(\mathbf{r})$ [17],

$$\mathbf{A}\Phi + \mathbf{B}(\delta)\Phi = \mathbf{S}, \quad (5)$$

where $\delta = \{\delta_1, \delta_2, \cdots, \delta_N\}$ is the vector of nodal abnormal absorption variations $\delta\mu_a(\mathbf{r})$, and the components of the matrix $\mathbf{A}$ is computed as

$$a_{ij} = \int_\Omega D(\mathbf{r}) \nabla \varphi_i(\mathbf{r}) \cdot \nabla \varphi_j(\mathbf{r}) d\mathbf{r} + \int_\Omega \mu_a^0(\mathbf{r}) \varphi_i(\mathbf{r}) \varphi_j(\mathbf{r}) d\mathbf{r} + \int_{\partial\Omega} \varphi_i(\mathbf{r}) \varphi_j(\mathbf{r}) / 2\alpha \, d\mathbf{r}, \quad i, j = 1, 2, \cdots, N, \quad (6)$$

and the components of the matrix $\mathbf{B}$ is computed as

$$b_{ij} = \sum_{k=1}^N \delta_k \int_\Omega \varphi_k(\mathbf{r}) \varphi_i(\mathbf{r}) \varphi_j(\mathbf{r}) d\mathbf{r}, \quad i, j = 1, 2, \cdots, N, \quad (7)$$

where $\varphi_i(\mathbf{r}) (i = 1, 2, \cdots, N)$ are the shape functions. Since the matrices $\mathbf{A}$ and $\mathbf{B}$ in Eq. (5) are positive definite, we have

$$\Phi = \left( \mathbf{A} + \mathbf{B}(\delta) \right)^{-1} \mathbf{S}. \quad (8)$$

To simplify the computational complexity, Eq. (8) can be linearized to establish the linear relation between the measured photon fluence rate $\Phi(\Gamma)$ and abnormal absorption variations $\delta$,

$$\Phi_0(\Gamma) - \Phi(\Gamma) = \mathbf{A}^{-1}(\Gamma)\mathbf{B}(\Phi_0)\delta, \quad (9)$$

where $\Gamma$ is a boundary nodal set on the finite element mesh corresponding to the measurable photon fluence rates on upper surface and lower surface of the breast, $\Phi_0 = \mathbf{A}^{-1}\mathbf{S}$ is the photon fluence rate without absorption perturbation. To find a stable solution of Eq. (9), we introduce the following optimization model,

$$\begin{cases} [\delta_{i_k}^*, p_k^*] = \arg\min \left\| \mathbf{A}^{-1}(\Gamma)\mathbf{B}(\Phi_0)\delta - \left(\Phi_0(\Gamma) - \Phi(\Gamma)\right) \right\| \\ \delta_{i_k} > 0, \ k = 1, 2, \dots, n \\ \delta_j = 0, \ j \notin \{ i_1, i_2, \dots, i_n \} \end{cases}, \quad (10)$$

where the vector $\delta$ only includes $n$ positive nonzero values $\left\{ \delta_{i_1}, \delta_{i_2}, \cdots, \delta_{i_n} \right\}$, which represents the abnormal absorption variations at isolated locations determined by the corresponding nodal indices $\{ p_1, p_2, \cdots, p_n \}$, and $n$ is a known positive integer which can be estimated as prior knowledge. For breast imaging, $n$ represents the number of tumors in the breast. Generally it is reasonable to assume that $n$ is less than 3 for the early stage breast tumor detection, helping to achieve a high computational efficiency of the optimization model (10). Each nodal index $p_k (k = 1, 2, \cdots, n)$ corresponds to the central position of a tumor in the breast.

Eq. (10) are a NP (nondeterministic polynomial time) problem and cannot be efficiently solved using a popular gradient-based optimization method. Hence, a differential evolution (DE)-based heuristic reconstruction algorithm can be used to solve the optimization problem. DE is a powerful stochastic global optimization strategy, which regenerates a population through executing some simple arithmetic operations such as mutation, crossover and selection, which has demonstrated a superior converging behavior and a high precision. This DE method is very efficient especially for the small number $n$ of variables. In case of the unknown $n$, we can also heuristically determine the number of tumors by gradually increasing the number of positive variations and solve the optimization problem in Eq. (10). For each reconstructed image, if two or more positive abnormal absorption variations are so close that they cannot be isolated within a user-provided spatial resolution, they should be combined into single abnormal absorption variations. Generally speaking, the number of tumors can be iteratively and heuristically determined to interpret measured data effectively.

### D. Quantification of absorption variations

Based on central locations of tumors estimated by the reconstruction algorithm in above subsection, we can further quantify the intensity distribution of abnormal absorption variations for all the tumors. We set a sub-region $\Omega_k (k = 1, 2, \cdots, n)$ around the tumor central position with a radius related to the estimation of the abnormal absorption variations $\delta_{i_k}^* (k = 1, 2, \cdots, n)$ to form a feasible region $\overline{\Omega} = \bigcup_{k=1}^{n} \Omega_k$ of abnormal absorption variations for image reconstruction. From the feasible region $\overline{\Omega}$, we perform image reconstruction using an algorithm for sparse equations and least squares (LSQR) [17, 18] based on Eq. (10) as follows:

$$\begin{cases} \boldsymbol{\delta}^* = \arg\min \left\| \mathbf{A}^{-1}(\Gamma)\mathbf{B}(\boldsymbol{\Phi_0})\boldsymbol{\delta} - (\boldsymbol{\Phi_0}(\Gamma) - \boldsymbol{\Phi}(\Gamma)) \right\| \\ \boldsymbol{\delta} \in \overline{\Omega}, \end{cases} \quad (11)$$

Using this methodology, the number of unknown variables is significantly reduced, and the image reconstruction of abnormal absorption variations in breast tissues can be performed effectively.

## III. NUMERICAL SIMULATION

To quantify the performance of proposed image reconstruction method, a digital breast phantom is applied for the numerical experiments. The breast phantom is developed from clinical dual-energy x-ray CT [20] (http://mcx.sourceforge.net/cgi-bin/index.cgi?DigiBreast). Two tetrahedral meshs, a finer one with 65196 tetrahedral elements and 14490 nodes for solving the optical forward model, and coarser one with 22799 tetrahedral elements and 4732 nodes for solving the inverse problem, are generated from the phantom. The phantom includes two different tumors in the shape of a 3D Gaussian-sphere to be expressed as $T(\mathbf{r}) = \exp\left(-\left\| \mathbf{r} - \mathbf{r_0} \right\|^2 / 2\sigma^2\right)$, their centers of two tumors are respectively set to $\mathbf{r_0} = (188, 41, 15)$ and $\mathbf{r_0} = (140, 49, 15)$, and the standard deviation of the Gaussian sphere is $\sigma = 1.7661 (mm)$. The simulations utilizes 20 continuous wave (CW) sources at 690 nm and 830 nm, 2622 CW detectors at upper surface, and 2473 CW detectors at lower surface. The absorption between 700nm-800nm is dominated by deoxy-hemoglobin while that between 800nm-900nm is dominated by oxy-hemoglobin, helping to recover the concentrations of deoxy- and oxy-hemoglobin.

**Table 1: Optical parameters**

| Tissue type | $C_{HbO2}$ | $C_{Hb}$ | $\mu_s'$ | |
|---|---|---|---|---|
| | | | 690nm | 830nm |
| Adipose | 13.84 | 4.81 | 0.851 | 0.713 |
| Fibroglandular | 18.96 | 6.47 | 0.925 | 0.775 |
| Malignant | 20.60 | 6.72 | 0.957 | 0.801 |

The fibroglandular tissue volume fraction map $C_f(r)$ is obtained from digital mammogram. In the phantom experiments, it is assumed that the primary breast tissue constituents are fibroglandular and adipose tissue, and the optical properties can be estimated by

$$\mu(r) = C_f(r)\mu_{fib} + C_a(r)\mu_{adi} \quad (12)$$

The optical coefficients of $\mu_{adi}$ and $\mu_{fib}$ are determined from adipose and fibroglandular optical properties, respectively,

based on the data in Table 1 [21]. $C_a(r)$ is the adipose tissue volume fraction at location $r$, which can be computed by $1 - C_f(r)$. The optical properties in the region of the tumor is estimated by

$$\mu(r) = \left(1 - C_t(r)\right)\left(C_f(r)\mu_{fib} + C_a(r)\mu_{adi}\right) \\ + C_t(r)\left(\mu_{adi} + \gamma(\mu_m - \mu_{adi})\right) \quad (13)$$

where $\mu_m$ is the optical coefficient of malignant tumor computed from the data in Table 1, and $\gamma$ expresses the tumor optical contrast [20] (here, $\gamma$ is set to 2).

A diffusion-equation based forward model was used to generate simulated CW optical measurements at 830 nm under various tumor settings. The diffusion equation was numerically solved on the forward mesh using the finite-element method. For each forward simulation, we added 5% Gaussian noise to the model output to simulate realistic measurement noise.

The image reconstruction algorithm in above Section was applied to identify the abnormal absorption regions from the generated simulated CW optical measurements data. Firstly, the central positions of two tumors were reconstructed at (188.35, 40.52, 14.31) and (139.18, 47.61, 15.29) based on optimization model (10). Then the quantification estimation of absorption perturbation was performed based optimization model (11) and the resultant relative errors in terms of absorption distribution were about 15%. We also conducted the image reconstruction with a compressive sensing (CS) regularization method to recovery the abnormal absorption variations from same dataset [17, 20]. These results show that the proposed method achieve accurate and stable image reconstruction clearly superior to the state of the art CS-based regularization reconstruction method [16, 21, 22], as shown in Fig. 3 (a-b).

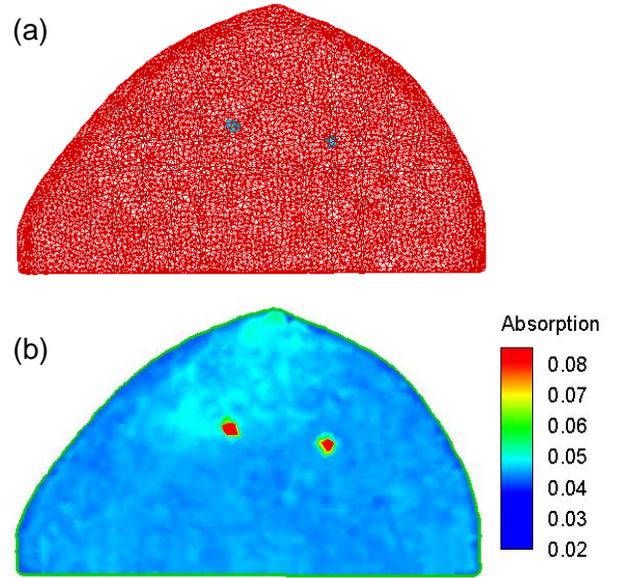

Figure 2. Breast phantom. (a) The 3D finite-element model with abnormal absorbing distribution, and (b) the true 2D slice of (a) at height=15 mm.

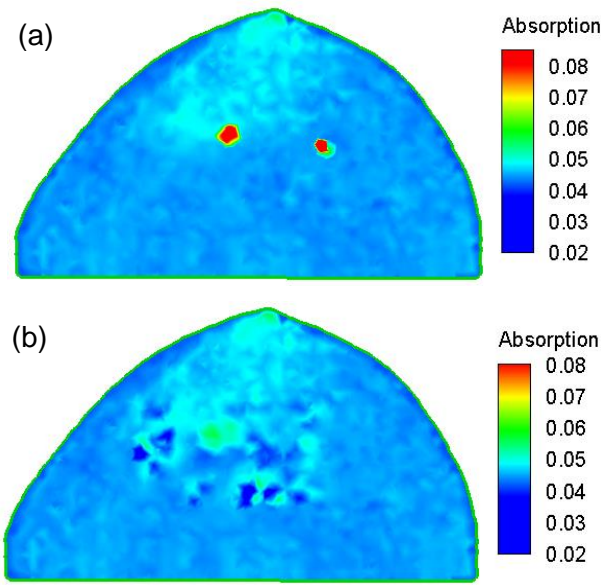

*Figure 3. Reconstruction of abnormal tissues. (a) The reconstructed 2D slice corresponding to the slice in Fig. 2(b) using the proposed method, and (b) the corresponding 2D slice reconstructed using compressive sensing (CS)-based regularization reconstruction method.*

## IV. DISCUSSIONS AND CONCLUSION

We have proposed a new optical tomographic imaging method for the detection of breast cancer. This method split the image reconstruction problem into the localization of abnormal tissues and quantification of absorption variations. The localization of abnormal tissues is performed based on a new well-posed optimization model, which can be solved via differential evolution optimization method to achieve a stable image reconstruction. For the early stage cancer detection, it is neither correct nor efficient to assume too many tumors. As far as the detection of a few numbers of tumors in a breast is concerned, the proposed image reconstruction method is very efficient to solve the optimization model of Eq. (10) for the determination of tumor central positions. Around the determined central positions of tumors, the quantification of abnormal absorption variations is then determined in localized regions of relatively small extents. Consequently, the number of unknown absorption variables can be greatly reduced to overcome the underdetermined nature of diffuse optical tomography, allowing for accurate and stable reconstruction of the abnormal absorption variations in the breast. The numerical simulations based on the realistic digital breast phantom have been performed to verify the feasibility of this method. The proposed method provides a practical and noninvasive tool for enhancing the accuracy of tumor diagnostics.

Our forward model is based on continuous wave (CW) excitation, which provide high signal-to-noise ratio of measured photon fluence rate. The proposed method can be directly extended to time-resolved/frequency-domain optical tomography, which provides more measurement information to enhance stability of image reconstruction. This proposed reconstruction method is based on the diffuse approximation model for high computational efficiency, which works well in breast tissues with weakly absorbing and highly scattering. It is also straightforward to extend our method to more accurate photon transport models, such as radiative transport equation (RTE) [14], phase approximation [15], or Monte Carlo simulation [22]. At same time, the graphic processing unit (GPU) can be used to achieve high performance computing to accelerate the computation of the photon transport models.